# Quantum sensing: Beyond the classical limits of precision[*]

Luiz Davidovich

*Institute for Quantum Science and Engineering, Texas A&M University, College Station, Texas 77843, USA*
*Universidade Federal do Rio de Janeiro, Rio de Janeiro, RJ 21941-972, Brazil*

Quantum sensors allow the estimation of parameters with precision higher than that obtained with classical strategies. Devices based on quantum physics have allowed the precise estimation of the gravitational field, the detailed imaging of the brain, the detection of gravitational-wave sources more than 400 million light years away, and an ever-increasing precision in the measurement of time. Quantum metrology, which is the conceptual framework that encompasses all these devices, is reviewed here, emphasizing recent results regarding noisy systems.

## 1. INTRODUCTION

Quantum sensing involves the use of quantum resources for the estimation of parameters characteristic of physical processes, with the aim of surpassing the precision obtained with classical strategies [1, 2]. It has become one of the most active areas of quantum information, with important theoretical developments and useful applications [3]. Quantum features, like entanglement and squeezing, have allowed the increase of precision of sensing devices, enabling the extension of the coverage of gravitational wave interferometers, with the use of squeezed light [4–6]; transportable gravimeters with resolution below $10^{-9}$ times the gravitational acceleration at the earth surface [7]; magnetometers that detect very weak magnetic fields, of the order of $10^{-11}$ Gauss [8, 9]; accelerometers with the potential of allowing GPS-free navigation [10]; and quantum-enhanced contrast and resolution in biological microscopy with squeezed light [11, 12].

The usual procedure, illustrated in Fig. 1, is to measure a probe that interacts with the process under investigation. Measuring the probe, after it undergoes a parameter-dependent dynamics, leads to an estimation of one or more parameters that characterize the physical process, through a function – an estimator – that maps an experimental data set to possible values of the parameters. There are four basic questions that one would like to answer: (i) How to define the precision of estimation?; (ii) How to get the precision from the experimental results?; (iii) What is the best initial state of the probe, in order to get the best precision?; and (iv) What is the best measurement procedure?

## 2. QUANTIFYING THE PRECISION IN PARAMETER ESTIMATION

Let us first discuss the precision of estimation in the classical case. Typically, the outcome of the measurement of the probe is described by a parameter-dependent probability distribution of the experimental results. The precision in the estimation of the parameter is related to the distinguishability between probability distributions corresponding to two nearby values of the parameter, which can quantified by the fidelity (assuming here a discrete probability distribution)

$$\Phi(X, X') \equiv \left[\sum_k \sqrt{P_k(X) P_k(X')}\right]^2, \qquad (1)$$

where $P_k(X)$ is the probability of getting an experimental value $k$ if the value of the parameter is $X$. As expected, the fidelity is equal to one if $X = X'$. For $X' = X + dX$, expansion in powers of *dX* yields

$$\lim_{dX \to 0} \Phi(X, X+dX) \to 1 - [F(X)/4](dX)^2, \text{ where } F(X) = 4\sum_k \left[\frac{d}{dX}\sqrt{P_k(X)}\right]^2. \qquad (2)$$

$F(X)$ is the *Fisher information*. From (2), one can see that $\sqrt{F(X)}/2$ is the speed of change of the fidelity, with respect to changes in the parameter $X$. For continuous distributions, $P_k(X)$ is replaced by the probability density $p(\xi, X)$ of getting an experimental result between $\xi$ and $\xi + d\xi$ if the value of the parameter is $X$.

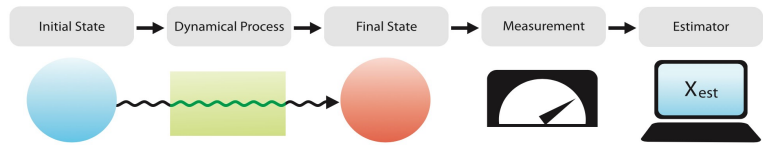

FIG. 1. Usual procedure for estimation of parameters. Measuring a probe, after it interacts with the system under investigation, leads to an estimation of one or more parameters that characterize the physical process, through a function – an estimator – that maps an experimental data set to possible values of the parameters.

---

One should expect that the larger the speed of change, the better the precision of the estimation of $X$. For unbiased estimations, the average of the estimator over a large number of realizations of the measurement coincides with the true value of the parameter. In this case, the precision of the estimation may be quantified by the standard deviation of the measured values of the parameter with respect to the average: $\Delta X = \sqrt{\langle X^2 \rangle - \langle X \rangle^2}$. Within the classical framework, a lower bound for the variance $(\Delta X)^2$ was obtained by Cramér and Rao [13], and shown by Fisher [14] to be attainable when the distribution of experimental values $\xi$ is Gaussian, or when the number of repetitions of the measurement is much larger than one. The Cramér-Rao bound expresses the standard deviation in terms of the Fisher information:

$$\Delta X \geq 1/\sqrt{\mathcal{N} F(X)}, \qquad (3)$$

where $\mathcal{N}$ is the number of independent measurements.

In order to generalize these results to quantum mechanics, one considers now the distinguishability of two quantum states, $|\psi(X)\rangle$ and $|\psi(X + dX)\rangle$, the corresponding fidelity being now defined by $|\langle \psi(X)|\psi(X + dX)\rangle|^2$. Then, expanding this expression in powers of $dX$, one gets

$$\lim_{dX \to 0} |\langle \psi(X)|\psi(X+dX)\rangle|^2 \to 1 - [\mathcal{F}_Q(X)/4](dX)^2, \qquad (4)$$

where $\mathcal{F}_Q(X)$ is the *quantum Fisher information* corresponding to the parameter $X$, which can therefore be expressed, according to (4), in terms of $|\psi(X)\rangle$ and its derivatives. Thus, similarly with the classical situation, the quantum Fisher information measures the speed of variation of the fidelity between quantum states with respect to changes in the parameter $X$. One should note, however, that while (2) depends on a specific measurement, characterized by the probability distribution $P_k(X)$, this is not the case for the quantum Fisher information, which is expressed in terms of the state $|\psi(X)\rangle$, and not on a probability distribution for a specific measurement. In fact, $\mathcal{F}_Q(X)$ corresponds to the best possible measurement made on the probe [1, 2], the one that maximizes the Fisher information, as given in (2), yielding the quantum bound on the precision for unbiased estimators: $\Delta X \geq 1/\sqrt{\mathcal{N} \mathcal{F}_Q(X)}$.

## 3. PARAMETER ESTIMATION FOR NOISELESS SYSTEMS

For noiseless systems and pure initial state of the probe, so that the state of the outgoing probe is $|\psi(X)\rangle = \boldsymbol{U}(X)|\psi(0)\rangle$, where $\boldsymbol{U}(X)$ is a unitary operator describing the parameter-dependent evolution of the probe, a simple analytical expression can be obtained, from (4), for the quantum Fisher information (boldface is used for operators):

$$\mathcal{F}_Q(X) = 4\langle (\Delta \boldsymbol{O})^2 \rangle_0 = 4\langle \psi(0)|[\boldsymbol{O}(X) - \langle \boldsymbol{O}(X)\rangle_0]^2|\psi(0)\rangle, \qquad (5)$$

where the index 0 specifies that the average is in the initial state of the probe, and the operator $\boldsymbol{O}(X)$ is defined as

$$\boldsymbol{O}(X) \equiv i[d\boldsymbol{U}(X)/dX]\boldsymbol{U}(X). \qquad (6)$$

In the special case when $\boldsymbol{U}(X) = exp(i\boldsymbol{G}X)$, with the operator $\boldsymbol{G}$, independent of $X$, being the *generator* of the unitary transformation $\boldsymbol{U}(X)$, one has $\boldsymbol{O} = \boldsymbol{G}$ and the quantum Cramér-Rao relation becomes

$$\Delta X \geq 1/(\sqrt{\mathcal{N}} \Delta G), \qquad (7)$$

with $\Delta \boldsymbol{G} \equiv \sqrt{\langle (\boldsymbol{G} - \langle \boldsymbol{G}\rangle_0)^2\rangle}$. This expression shows that a strategy for increasing the precision consists in increasing the variance of the operator $\boldsymbol{G}$ in the initial state of the probe.

If $X$ is taken as an evolution time of the probe, then $\boldsymbol{G}$ is the Hamiltonian $\boldsymbol{H}$ that generates this evolution, and, for $\mathcal{N} = 1$, one has, from (7), $\Delta \boldsymbol{H} \Delta t \geq 1$, which is the energy-time uncertainty relation, a consequence therefore of the Cramér-Rao relation, not of the Heisenberg uncertainty relation (which could not be, since time is a parameter, not an operator).

### 3.1. Example: Phase estimation in optical interferometry

Figure 2 illustrates a Mach-Zender interferometer, used to estimate the phase shift of a light beam, due to the presence of a transparent sample in one of the arms of the interferometer. This would allow the estimation of the refractive index of the sample. The probe, in this case, is the light field. The phase shift operator is $\boldsymbol{U}(\theta) = \exp(i\theta \boldsymbol{n})$, where $\boldsymbol{n} = \boldsymbol{a}^\dagger \boldsymbol{a}$ is the number operator, and $\boldsymbol{a}^\dagger$, $\boldsymbol{a}$ are the photon creation and annihilation operators. Fock states $|n\rangle$, which have a well-defined number of photons, are eigenstates of $\boldsymbol{n}$: $\boldsymbol{n}|n\rangle = n|n\rangle$, so that $\boldsymbol{U}(\theta)|n\rangle = exp(i\theta n)|n\rangle$. Also, for a coherent state $|\alpha\rangle$ — an eigenstate of the annihilation operator $\boldsymbol{a}$, with $\boldsymbol{a}|\alpha\rangle = \alpha|\alpha\rangle$ — one has $\boldsymbol{U}(\theta)|\alpha\rangle = |\alpha \exp(i\theta)\rangle$, which justifies defining $\boldsymbol{U}(\theta)$ as a phase-shift operator, with generator $\boldsymbol{n}$.

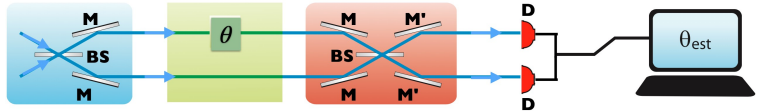

FIG. 2. Mach-Zender interferometer, used to estimate the phase shift θ induced on a light beam by a transparent sample. The interferometer is composed of two beam splitters (**BS**) and four mirrors (**M**), plus two external mirrors (**M'**) that deflect the output light beams towards the detectors (**D**). Detection of the outgoing beams allows the estimation of the phase shift.

From (7), it follows then (taking $\mathcal{N} = 1$) that $\Delta\theta\Delta n \geq 1/2$, with $\Delta n$ being the standard deviation of the number of photons in the light beam that crosses the sample. This is a phase-number uncertainty relation, a consequence of the Cramér-Rao relation, which does not require therefore the definition of a phase operator: here, phase is considered as a parameter.

For a coherent state, with a Poissonian distribution of the photon numbers, $\Delta n = \sqrt{\langle n \rangle}$, where $\langle n \rangle$ is the average number of photons in the state, so that $\Delta\theta \geq 1/\left(2\sqrt{\langle n \rangle}\right)$. This lower bound is known as the *standard limit* for phase estimation. A better precision can be obtained if the photons in the interferometer are described by the entangled NOON state [15] $|\psi(N)\rangle = (|N, 0\rangle + |0, N\rangle)/\sqrt{2}$, where $|N, 0\rangle$ has $N$ photons in the upper arm of the interferometer, so that $|\psi(N)\rangle \to [exp(i\theta N)|N, 0\rangle + |0, N\rangle]/\sqrt{2}$. Then $\Delta n = N/2$, implying that $\Delta\theta \geq 1/N$, so precision is increased for the same number of photons, as long as $N > 1$. This is the so-called *Heisenberg limit* for phase estimation.

This example shows how quantum mechanical properties like entanglement can help to increase the precision in parameter estimation. However, the NOON states are difficult to produce, and furthermore they are very sensitive to noise: The loss of a single photon by a NOON state destroys its entanglement. The final state is either $|N - 1, 0\rangle$ or $|0, N - 1\rangle$, with the same probability, and neither of them is useful for phase estimation. This sensitivity is typical of entangled states [16–19]. Therefore, under the omnipresence of noise, new methods are needed in order to find the best achievable precision and the corresponding best states of the probe.

## 4. QUANTUM METROLOGY OF NOISY SYSTEMS

As shown in the previous sessions, for noiseless quantum processes, with unitary evolution of probe dynamics, and unbiased estimators, simple expressions are obtained for the quantum Cramér-Rao bound. This is not so, however, for open systems, that is, systems in the presence of an environment. Exact solutions can be found for Gaussian states of the probe [20–22], or for one or two qubits, but it is not possible, in general, to find analytical solutions. Upper bounds for the quantum Fisher information can be found through purification of the non–unitary dynamics, by adding an ad hoc environment, such that the dynamics of the enlarged system is unitary and the reduced dynamics, obtained by tracing out the added environment, coincides with the original dynamics of the system [23–26]. In [23] it is shown that it is always possible to find a purification so that the upper bound coincides with the quantum Fisher information of the noisy system. This was achieved in [25], where it was derived an analytic expression for the best possible precision in the estimation of a force applied to a damped harmonic oscillator, in contact with a thermal bath. Even when this ideal upper bound is not found, very good analytic approximations are obtained for the precision [23, 24, 26].

## 5. ESTIMATION OF ABSORPTION PARAMETERS

An important application of the framework of quantum metrology for noisy systems is the estimation of absorption of light by a sample. The absorption coefficient $\gamma$ is defined in terms of the output and input average photon numbers (proportional to the respective intensities) of the probing light beam by $\langle n \rangle_{out} = (1 - \gamma)\langle n \rangle_{in}$ — see Fig. 3(a). It is known that maximum precision is obtained by using Fock states and photon-counting detection [27–29]. Fock states are however hard to produce, except for small photon numbers. An alternative strategy was proposed in [30]: The absorption medium is placed between two optical parametric amplifiers (OPAs). The first one, with vacuum input, produces a two-mode vacuum squeezed state, with average photon number $\langle n \rangle_{in}$ in each mode, one of the two beams probing the medium and the other playing the role of an ancilla. The second OPA reverses the squeezing transformation, so that, in the absence of the absorption medium, there is no outgoing field.

Detection of the output photons leads to the estimation of the photon-loss coefficient $\gamma$. Fig. 3(b) illustrates this configuration, which defines a SU(1,1) interferometer [31, 32]. Since the two-photon squeezed state is Gaussian, the quantum Fisher information can be obtained exactly. Interestingly, it coincides with the one corresponding to a Fock state, with photon number replaced by the average number of photons in each beam.

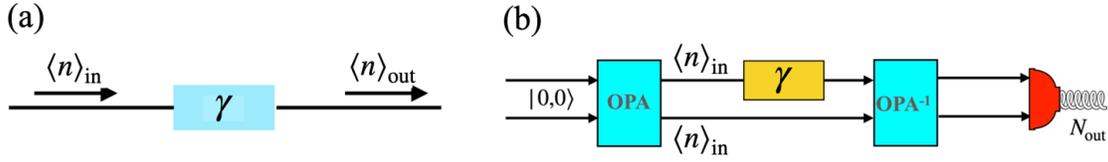

FIG. 3. Setups for (a) single-mode and (b) squeezed-vacuum time-reversal estimation of absorption. In (b), the sample is placed between two optical parametric amplifiers (OPAs). The first OPA produces a two-mode squeezed state. Each of the two modes has an incoming average photon number $\langle n \rangle_{in}$. The second OPA reverses the squeezing transformation.

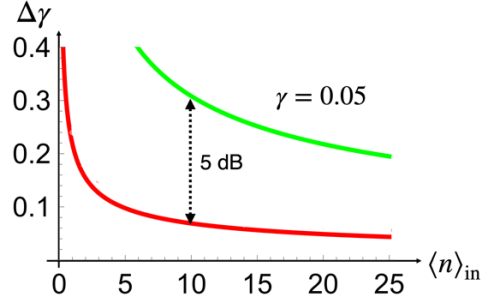

FIG. 4. Uncertainty in the estimation of the absorption constant, for $\gamma = 0.05$. Comparison between the bound for the uncertainty $\Delta\gamma$ obtained from the quantum Fisher information for probe plus ancilla corresponding to two modes of an incoming bimodal squeezed state (red line) and the standard limit, obtained for a single-mode coherent state testing the sample, with the same average photon number as each mode of the bimodal squeezed state (green line).

The uncertainty in the estimation of $\gamma$ can be obtained through the relation

$$\Delta\gamma = \Delta N_{out}/(d\langle N\rangle_{out}/d\gamma) , \qquad (8)$$

where $\Delta N_{out}$ is the standard deviation of the output photon-number distribution after the second OPA, and $\langle N \rangle_{out}$ the corresponding average photon number. For small absorption, the precision obtained from (8) is practically indistinguishable from the one given by the quantum Fisher information, and much better than the one obtained with a single-mode configuration, with a coherent state as input, which corresponds to the standard limit in this case. Fig. 4 compares the uncertainty in the estimation of $\gamma$ for the SU(1,1) configuration with the standard limit. The quantum advantage is evident: for the probe plus ancilla setup, the increase in precision from the standard limit is about 5 dB for $\langle n \rangle_{in} = 10$. As the absorption increases, so that $\gamma \to 1$, the precision for the bi-modal squeezed state approaches the standard limit, an example of the decoherence-induced classical limit of quantum physics [33, 34].

## 6. ACKNOWLEDGMENTS

It is a pleasure to acknowledge the collaboration on quantum metrology with students and colleagues at the Texas A&M University and at the Federal University of Rio de Janeiro: Girish S. Agarwal, Bruno M. Escher, Camille L. Latune, Ruynet L. de Matos Filho, Alvaro H. Pimentel, Márcio M. Taddei, Fabricio Toscano, Stephen P. Walborn, Jiaxuan Wang, and Nicim Zagury. I acknowledge the support of the Brazilian agencies CNPq, CAPES, and the Rio de Janeiro State Foundation for Research Support (FAPERJ).